# The MCS Model of the Superconductivity in Hole-Doped Cuprates: an Experimental Evidence

A. Mourachkine

*Université Libre de Bruxelles, Service de Physique des Solides, CP233, Boulevard du Triomphe, B-1050 Brussels, Belgium*

We present experimental data obtained by different techniques. Some of them are not easy to explain by any theoretical model presented in the literature. However, a MCS model of the superconductivity in hole-doped cuprates, proposed in our previous work, is capable to explain them. Thus, we present an evidence in favor of the MCS model.

## 1. INTRODUCTION

The mechanism of superconductivity (SC) in high-$T_c$ superconductors (HTSC) [1] remains an open question. There is a clear microscopic difference between the BCS superconductors and HTSC, namely, that they have a different origin and that different criteria are required for the HTSC than for the conventional SCs. As for the bulk characteristics, the controlling factor in cuprates seems only to be the hole density in $CuO_2$ planes. The definitive confirmation of the predominant $d_{x^2-y^2}$ (hereafter, d-wave) character of the SC in hole-doped cuprates is a great advance in recent years [2]. At the same time, the order parameter (OP) in the electron-doped cuprate, $Nd_{2-x}Ce_xCuO_4$ (NCCO), has a s-wave symmetry. The normal-state properties of cuprates are unique, physicists never dealt with such bizarre normal-state properties before. As a consequence of all these facts, there are too many models of HTSC, some of which are pure theoretical. At the same time, there are many experimental data which can not be explained by none of these models.

There is clear evidence for charge stripe formation in $La_{1.6-x}Nd_{0.4}Sr_xCuO_4$ [3] and $YBa_2Cu_3O_{6.6}$ (YBCO) [4]. Holes induced in $CuO_2$ planes segregate into periodically-spaced stripes that separate antifferomagnetic (AF) isolating domains. By using this experimental fact, Emery, Kivelson and Zachar [5] presented a stripe model of underdoped HTSC. The main feature of the model is a spinon SC. Spinons are neutral fermionic excitations which appear in one-dimensional (1D) physics. The spinon SC occurs on charge stripes in $CuO_2$ planes at $T_{pair} > T_c$. Spinons create pairs on stripes because of the existence of a large spin-gap in AF domains due to AF correlations [5]. The coherent state of the spinon SC is established at $T_c$ by the Josephson coupling between charge stripes carrying the spinon SC. Figure 1 shows *schematically* a snapshot of the stripe model.

In our previous works [6,7], we have presented an evidence of the existence of two SC gaps and magnetic origin of the coherent gap in hole-doped cuprates. An attempt to find a theoretical model among that presented in the literature in order to explain the data is failed. However, we found that the stripe model may fit the data after a serious modification, namely, if instead of the Josephson coupling between stripes we introduce the second SC condensate with the magnetic coupling. So, we proposed a MS (Magnetic polarons - Spinons) model [6], and later a MCS (Magnetic Coupling of Stripes) model [8] in which interstripe coupling occurs directly due to spin fluctuations.

In this paper, we present experimental data obtained by different techniques. Some of the data are obtained in the presence of a magnetic field, and it is not easy to explain them by any theoretical model presented in the literature. However, the MCS model is capable to explain them. Thus, we present an evidence in favor of the MCS model.

The structure of the paper is as follows. The MCS model is briefly described in Section 2. In Section 3, we present experimental data obtained by different techniques in the presence of a magnetic field. The direct evidence in favor of the MCS and MS models is presented in Sections 4. In Section 5, we focus attention on the analogy and difference between the SC in heavy-fermion compound $UPd_2Al_3$ and in hole-doped cuprates. The final discussion is presented in Section 6. The conclusions can be found in Section 7.

## 2. THE MCS MODEL

In this Section, we describe briefly the MCS model.

Emery and Kivelson [9] pointed out that a finite density of holes in cuprates forms self-organized structures, designed to lower the zero-point kinetic energy. This is accomplished in three stages: a) the formation of charge inhomogeneity (stripes), b) the creation of local spin pairs, and c) the establishment of a phase-coherent state. In general, we agree with such description of the SC in cuprates. However, the third step in their model [5], *i.e.* the coherent state, is established at $T_c$ by the Josephson coupling between charge stripes. We showed that the coherent-SC gap has the magnetic origin [6]. Thus, this fact is in contradiction with Emery, Kivelson and Zachar's model. So, we proposed that *fluctuating* stripes couple to each other due to spin fluctuations which propagate into AF domains [8]. We call this model as the Magnetic Coupling of Stripes (MCS) model. In frameworks of the MCS model, the SC has two different mechanisms: along charged stripes for coupling and perpendicular to stripes in order to establish the coherent state (see Fig. 1). Thus, there are, at least, two different SC gaps in cuprates. Figure 2 shows a phase diagram of $Bi_2Sr_2CaCu_2O_{8+x}$ (Bi2212) [8]. In Fig. 2, the coherent gap, $\Delta_c$, scales with $T_c$ as $2\Delta_c/k_BT_c = 5.3$ [6,8]. The spinon gap, $\Delta_s$, and $T_c$ do not relate to each other. As a consequence of charge inhomogeneity in cuprates, carriers exhibit different properties in different directions: *fermionic* along charged stripes and *polaronic* perpendicular to stripes.

Before the MCS model, we introduced the MS (Magnetic polarons - Spinons) model [6]. In frameworks of the MS scenario, there are two SC condensates in $CuO_2$ planes: one on charge stripes like in the MCS model, and the second condensate between charge stripes, *i.e.* in AF domains (see Fig. 1). The quasiparticles in AF domains are magnetic polarons. The magnetic polaron can be pictured as the electron and a spin polarization around it [10]. Magnetic polarons couple to each other due to spin fluctuations and establish the coherent state at $T_c$. The coherent state of the spinon SC which exists above $T_c$ is established via the magnetic-polaron SC. Thus, in latter scenario, the fluctuations of charge stripes are not so important.

In both scenarios, the SC gap with the magnetic origin is predominant and has the d-wave symmetry [7]. The spinon-SC gap along charge stripes has a s-wave or mixed (s+$d_{xy}$) symmetry [6]. It is possible that, in hole-doped cuprates, there exists a small SC gap having a g-wave symmetry, which originates from the magnetic coupling mechanism [6].

There are two special cases of the SC in cuprates, namely, in NCCO and YBCO. It is well known that, in the electron-doped NCCO cuprate, the OP has a s-wave symmetry. This means that the SC with the magnetic pairing due to spin fluctuations is absent in NCCO. We have proposed that the spinon SC on charge stripes in NCCO is similar to that in hole-doped cuprates. However, in frameworks of the MCS model, the interstripe coupling occurs due to phonons, or, in frameworks of the MS model, quasiparticles in AF domains are usual lattice polarons which condense into the Cooper pairs at $T_c$. So, the MCS model of the SC in hole-doped cuprates is to be called as a PCS (Phonon Coupling of Stripes) model for the case of NCCO. The SC in all cuprates occurs into $CuO_2$ planes. The only cuprate which has Cu-O chains is YBCO. In YBCO, there is an additional SC gap on chains which has to mimic the behavior of the coherent gap since the SC on chains is induced due to the proximity effect [11]. The chain SC occurs also due to pairing of spinons since chains are one-dimensional. These phenomena await future investigations.

From our previous work [8], the scenario in frameworks of the MCS model is more likely than the MS scenario. Nevertheless, further, we will use a general abbreviation for the two models, the M(C)S models, since the MCS and MS models can not be distinguished from the interpretation of data presented in following Sections.

## 3. DATA OBTAINED IN A MAGNETIC FIELD

The SC and a magnetic field avoid each other: $B = 0$ inside a superconductor, where $B$ is the magnetic induction. Thus, the study of the interplay between a SC and a magnetic field is very useful for the understanding of the mechanism of a SC. In this Section, we present experimental data obtained in the presence of an external magnetic field. After the presentation of each set of the data, we will give comments in frameworks of the M(C)S models.

Very recently, in NMR measurements on near optimally doped YBCO, Gorny *et al.* [12] found that a magnetic field of 14.8 Tesla shifts $T_c$ down by 8 K while the pseudogap above $T_c$ remains unaffected (as measured by $1/(T_1T)$). In frameworks of the M(C)S models, this means that a magnetic field suppresses the magnetic coupling mechanism either for stripes or magnetic polarons, while the value of the magnetic field of 14.8 Tesla is not enough to destroy the Cooper pairs which exist above $T_c$ on stripes.

Rener *et. al.* [13] observed on Bi2212 by STM a large gap into the vortex core at low temperature. Figure 3 shows three types of tunneling spectra obtained on under- and overdoped Bi2212 single crystals: at the center of a vortex core at 4.2 K; between vortices at 4.2 K, and above $T_c$ in zero magnetic field. One can see in Fig. 3 that the gaps at the center of the cores are very similar to the pseudogaps measured above $T_c$. In Fig. 3, one can mention also that the magnitudes of gaps measured into vortex cores are larger than that outside the vortex cores. In frameworks of the M(C)S models, this means that inside the vortex cores, the SC with the magnetic pairing is completely suppressed while the spinon SC is unaffected, and it is similar to the case above $T_c$. It is logic that the upper critical magnetic field along stripes is higher than the upper critical magnetic field of the magnetic coupling, $H_{c2,\Delta s} > H_{c2,\Delta c}$ since $\Delta_s > \Delta_c$, always (see Fig. 2). The asymmetry of the spectra measured into vortex cores and shown in Fig. 3 we attribute to interactions of the dc current with spin channel [14].

In tunneling measurements on overdoped Bi2212 single crystals in SC-insulator-SC (SIS) junctions, Vedeneev *et al.* [15] found that tunneling spectra measured at 4.2 K, which are shown in Fig. 4, are not affected too much by a magnetic field of 20 Tesla. However, independently from the direction of the magnetic field ($B \parallel$ c-axis or $B \perp$ c), the magnitude of tunneling gap in a magnetic field of 20 Tesla becomes slightly larger. In frameworks of the M(C)S models, this means that a magnetic field of 20 Tesla at 4.2 K suppresses *partially* the SC gap having the magnetic origin while the spinon-SC gap is not affected. These data are very similar to the data obtained between vortices by STM [13].

The same effects have been observed in YBCO [16,17,13]. Figure 5 shows tunneling spectra measured on near optimally doped YBCO in $B = 0$ and $B = 8$ T ($B \parallel$ c) at 10 K [16]. First of all, we focus attention on the conductance curve in zero magnetic field (solid line). In Fig. 5, one can see that there are three prominent features in $G(V)$, namely, symmetric peaks located at $\pm 19$ mV, asymmetric peaks at $\pm 36$ mV, and a subgap at $\pm 4$-5 mV [16]. In fact, we will see further that they are typical features in YBCO. In frameworks of the M(C)S models, we attribute the peaks located at $\pm 19$ mV to the coherent-SC gap having the magnetic origin and peaks at $\pm 36$ mV to the spinon-SC gap (see Fig. 2). The subgap can be attributed to a SC gap on chains. However, Krezin and Wolf showed that, theoretically, an induced SC gap on chains is larger than that in $CuO_2$ planes. In a magnetic field of 8 Tesla, these three gaps are affected differently. The peaks located at $\pm 19$ mV are almost suppressed. The peaks at $\pm 36$ and $\pm 4$-5 mV are practically unchanged. This result is in a good agreement with the M(C)S models. A magnetic field of 8 Tesla is not enough to destroy spinon pairs along stripes. The peaks at $\pm 4$-5 mV are unaffected since the induced SC on chains occurs due to pairing of spinons. The spinon SC is more difficult to suppress than the SC with the magnetic pairing. In Fig. 5, the peaks at ± 36 have a tendency to become symmetric in a magnetic field.

On near optimally doped YBCO, Maggio-Aprile *et al*. [17] observed at 4.2 two gaps into vortex cores: a large gap like in Bi2212 and small gap. Figure 6 shows a tunneling spectrum obtained in zero magnetic field and average spectrum at the center of a vortex core. The tunneling spectrum in zero magnetic field has the same three features as the spectrum shown in Fig. 5, namely, the peaks located at ± 20 and ± 30 mV and the subgap at low bias (~ 5-6 meV). The spectrum at the center of a vortex core (lower curve in Fig. 6) displays clearly two gaps which coincide with the largest gap and subgap shown on upper spectrum in Fig. 6. The small gap of 5.5 meV measured in the center of the vortex core we attribute to the spinon SC on chains, induced along chains from outside the vortex core. The peaks located at ± 20 mV on the upper spectrum in Fig. 6 are completely suppressed in the center of a vortex core (lower curve in Fig. 6). The results of these measurements [17] are almost identical to the data shown in Fig. 5 [16]. Thus, the explications in frameworks of the M(C)S models for the data shown in Fig. 6 can be found in the previous paragraph.

In microwave measurements of the complex surface impedance $Z_S = R_S + iX_S$ *in the mixed state* on slightly overdoped Bi2212 single crystals, Hanaguri *et al*. [18] found that $X_S$ which is proportional to the real part of the effective penetration depth increases while there was little changes in $R_S$. The value of the penetration depth is a characteristic of the coherent state while $R_S$ is a characteristic of quasiparticle transport. Thus, in frameworks of the M(C)S models, the experimental data can be explained by the fact that, in the mixed state, the SC gap having the magnetic origin is seriously damaged while the spinon-SC gap is unaffected. The changes in $R_S$ are small since the Cooper pairs on stripes are mainly responsible for the value of $R_S$.

It is obvious that the results obtained on Bi2212 [13,15] and on YBCO [16,17] are very similar with the exception of the SC on chains, of course. On the other hand, measurements show that this picture is different than that in NCCO. In Section 2, we speculated that the coherent state of the spinon SC in NCCO occurs at $T_c$ either due to phonons or due to lattice bipolarons formed in AF domains. Indeed, it can be the case. The change in resistivity transition induced by a magnetic field in NCCO exhibits a different behavior than that in other cuprates [19]. Usually, in hole-doped cuprates, applying a field causes profound broadening of the resistivity transition (from the M(C)S models, it is clear why), while in NCCO, applying a field ($H \parallel c$) causes the parallel shift of transition curves to lower temperatures [19]. Thus, this picture resembles the case for conventional type-II SCs. However, at the same time, the upper critical magnetic field, $H_{c2}$, in NCCO is not drastically different than in other one-layer cuprates [19]. This picture is completely in an agreement with the PCS model (see Section 2). The value of $H_{c2}$ in NCCO is not different than that in other one-layer cuprates since the spinon SC on stripes is similar in all cuprates including

NCCO. The main difference between the two systems is the mechanism of establishment of the coherent state which occurs either due to spin fluctuations or phonons.

In summary, all data presented in this Section can be explained naturally by the M(C)S models. From these data along, it is not possible to give a preference to one of the two models, to the MCS or MS model. It seems that the PCS model of the SC in NCCO looks also plausible.

## 4. OTHER EXPERIMENTAL DATA

In this Section, we present a few pieces of direct evidence in favor of the M(C)S models.

On underdoped Bi2212, Corson *et al.* [20] found the presence of the Cooper pairs above $T_c$ in measurements of high-frequency conductivity that track the phase-correlations time in the normal state of Bi2212. This result is in a good agreement with the M(C)S models.

The symmetries of the two SC gaps are not very important to discuss the M(C)S models. However, theoretically, the magnetic origin of the coherent gap implies the d-wave symmetry of the $\Delta_c$, and it is in a good agreement with the experiment [21,7]. In our previous works [6,7], we showed that the spinon-SC gap has either a s-wave or a mixed ($s+d_{xy}$) symmetry. Thus, the spinon-SC gap has entirely or partially a s-wave symmetry. Rossel *et al.* [22] presented an experimental evidence of the co-existence of predominant d-wave component with an admixture of a s-wave component in $Tl_2Ba_2CuO_{6+x}$ (Tl2201) cuprate. In YBCO, a s-wave component has to be present too, otherwise, it is not possible to explain the Fraunhofer pattern of the Josephson current, $I_c(B)$ observed in Pb/insulator/YBCO junctions [23].

With the exception of our work [8], there is an evidence in favor of the MCS model rather than in favor of the MS model. In magnetoresistance measurements on undoped YBCO, Ando, Lavrov and Segawa [24] found the existence of charge stripes in YBCO. They observed that stripes can be directed with external magnetic field, thus, stripes should have local ferromagnetic moment. This fact contradicts to Tranguada's model [25] in which spin orientations across stripes are opposite. On the other hand, this is an evidence in favor of the MCS model. Very recently, Tranquada [26] excepted the fact that so-called resonance peak observed by inelastic neutron scattering (INS) in YBCO and Bi2212 [27] originates from AF domains of $CuO_2$ planes and not from charge stripes.

Any SC gap in an applied magnetic field is entirely or partially suppressed. In frameworks of the M(C)S models, the spinon gap in cuprates is the last one to be suppressed by a magnetic field. From the dependence $H_{c2}(N)$, where N  1 is the number of $CuO_2$ planes, it is clear that the spinon gap is intimately related to magnetic interactions into $CuO_2$ planes. It is known that magnetic interactions in (N+1)-layer cuprates are stronger than in N-layer cuprates. The value of the upper magnetic field, $H_{c2}$, is varied from ~25 Tesla in one-layer cuprates [28] to 160 Tesla in four-layer cuprate [29]. In double-layers compounds YBCO and Bi2212, $H_{c2}(0)$ is around 90 Tesla [29]. This means that the value of the spinon-SC gap is related to magnetic correlations into

CuO$_2$ planes. This fact had been predicted in the stripe model [5], and it is in favor of the M(C)S models too.

From femtosecond time-resolved spectroscopy on near optimally doped YBCO, Stevens *et al.* [30] presented an evidence for the existence of two components: band-like and polaronic-like carriers in YBCO. Mihailovic and Müller [31] presented also strong indications for simultaneous presence of polaronic and fermionic carriers in mesoscopic areas in the CuO$_2$ planes below $T_c$. In frameworks of the M(C)S models, this fact is obvious. On the other hand, none of theoretical models can explain this fact, even, the stripe model [5].

In spite of experimental facts there is a skepticism into the HTSC community about the presence of two energy scales in SC cuprates. Therefore, we concentrate now on the presence of two SC gaps in cuprates. The idea of the presence of two SC gaps in cuprates is not new [11, 32]. Very recently, Deutscher [33] showed unambiguously the presence of two energy scales in hole-doped cuprates. These two energy scales match the two gaps shown in Fig. 2. Although, the spinon-SC gap is an excitation gap in his interpretation. On the other hand, Miyakawa *et al.* [33,34] showed from tunneling experiments that a tunneling gap with the maximum magnitude, which depends linearly on hole concentrations, is a SC gap. So, the only question is: are they real, these two SC gaps in cuprates? Further, we will demonstrate unambiguously the presence of two energy gaps in cuprates on the basis of tunneling measurements.

Figure 7 shows *typical* tunneling spectra obtained on (a) single-layer Tl2201 with $T_c \sim 91$ K [35]; (b) double-layer Bi2212 with $T_c = 89.5$ K [7], and (c) double-layer YBCO with $T_c = 89$ K (upper curve) [16] and with $T_c = 91$ K (lower curve) [17]. The spectra of Tl2201 and YBCO are obtained in SC-insulator-normal metal (SIN) junctions and the two spectra of Bi2212 are obtained in one SIS junction. It is important to note that (i) these three different cuprates have similar $T_c \sim$ 89 - 91 K and near optimally doping; (ii) the two spectra shown in Fig. 7(b) are obtained in *one* break junction, *i.e.* on same Bi2212 single crystal, and present the minimum and maximum gap magnitudes in a junction; and (iii) the data presented in Fig. 7 are *typical* for more than 100 junctions in each case.

The tunneling measurements on Tl2201 show definitely one energy gap with the magnitude of $\Delta = 20 - 22$ meV [35] and very specific shape of tunneling spectra (the spectrum A in Fig. 7). The measurements on YBCO show three energy gaps [16,17] which we discussed already in Section 3. We will discuss subgaps observed in spectra of YBCO (the spectra D and E in Fig. 7) and Bi2212 (the spectrum B in Fig. 7) in Section 6. At this moment, we concentrate on the two large energy scales in cuprates. From Fig. 7(c), the presence of two energy scales in YBCO, $\Delta_1 = 19\text{-}20$ meV and $\Delta_2 = 30\text{-}36$ meV, is obvious. Their magnitudes are in a good agreement with the phase diagram shown in Fig. 2, which has been obtained on Bi2212 [6,8]. It seems that, in YBCO, these two gaps, the $\Delta_c$ and $\Delta_s$, are difficult to separate in tunneling measurements. In Bi2212, they can be observed separately (the spectra B and C in Fig. 7) or together on same tunneling spectrum

(see, for example, a spectrum C in Fig. 1 of Ref. 7; Figs. 3 and 4(a) of Ref. 37, Fig. 3(b) of Ref. 38). The peaks in the spectra B and C shown in Fig. 7(b) belongs to the two different energy scales, namely, the $\Delta_c$ and $\Delta_s$. It is interesting that they have different shapes and different relative amplitudes: the amplitude of the $\Delta_c$ peaks is higher than the amplitude of the $\Delta_s$ peaks in approximately 1.5 times. In tunneling measurements by STM on near optimally doped Bi2212, Chang *et al*. [39] observed the *same* effect. Along 500 Å scan, they viewed the change in the gap magnitude between $\Delta_1 = 21$ meV and $\Delta_2 = 32$ meV, at that the maximum and minimum amplitudes of tunneling peaks have been observed when the gap magnitude was equal to $\Delta_1 = 21$ meV and $\Delta_2 = 32$ meV, respectively (see Figs. 2 and 3 of Ref. 39). The ratio between the maximum and minimum peak amplitudes in their measurements is equals to 1.3. In tunneling measurements on Bi2212, Renner and Ficher [37,40] observed *identical* effect along 500 Å scan. The ratio between the maximum and minimum peak amplitudes in their measurements is equals to 1.5. However, the smaller energy scale detected in the experiment was equal to 27 meV. We explain this small increase in the gap magnitude by the fact that the measurements have been performed in a magnetic field of 1.5 Tesla [37]. Thus, the presence of the two energy scales in Bi2212 have been reported already a few times. Moreover, the variation of the magnitude of tunneling gap, on average, between 21 meV and 37 meV have been observed also in angle resolved tunneling measurements [41]. All measurements on near optimally doped Bi2212 show consistent values of the two energy scales, $\Delta_1 = 20\text{-}23$ meV and $\Delta_2 = 32\text{-}38$ meV, which are in a good agreement with the similar values for YBCO and in Fig. 2. Even, the shapes of main tunneling peaks in the spectra A, B and E shown in Fig. 7, which are obtained on three different families of cuprates and correspond to the smaller energy scale, are very similar. The only question we still have to answer, why do tunneling measurements on Tl2201 show exclusively one SC gap, namely, the $\Delta_c$? However, there is no doubts that the second SC gap, the $\Delta_s$, is present in Tl2201, moreover, we know that it has a s-wave symmetry [22]. The question why it is not possible to detect it in tunneling measurements we will discuss in Section 6.

Despite the facts presented above, which clearly show the presence of two different energy scales in cuprates, one can still argue that these results are pure coincidence. We discuss further the temperature dependence of quasiparticle density of states (DOS) measured in cuprates, which can not cheat. The temperature dependence of quasiparticle DOS in Bi2212 and YBCO demonstrate clearly the presence of two OPs. Figure 8 shows different temperature dependencies of quasiparticle DOS, presented in the literature [7, 34, 42-44]. In Fig. 8, one can see that there are two distinct types of temperature dependencies of tunneling gap, $\Delta(T)$ [7]. One type of $\Delta(T)$ follows the BCS temperature dependence at low temperatures, and, at temperatures close to $T_c$, lies above the BCS dependence [34,42]. The SC gap corresponding to this type of $\Delta(T)$ has a tendency to evolve continuously into a normal-state quasiparticle gap. The second type of $\Delta(T)$ lies well below the BCS temperature dependence [7,43,44]. In Fig. 8, one can see that quasiparticle

DOS can "jump" from one type of $\Delta(T)$ to another type (the curve with open circles in Fig. 8). It was shown earlier that in two-band SC, the temperature dependence of the smaller OP appears below the BCS-temperature dependence [45,46]. This means that, in Bi2212 and YBCO, there are two OPs. In order to make easier to see it, In Fig. 9, we present separately temperature dependencies of two different gaps observed *simultaneously* on same spectrum of Bi2212 (Fig. 3(b) of Ref. 38). In Fig. 9, the curve which lies below the BCS dependence corresponds to the smaller gap, $\Delta_1 = 21.5$ meV, and the curve above the BCS dependence corresponds to the larger gap, $\Delta_2 = 35$ meV at 4.2 K [38]. The difference in the behavior of the two tunneling gaps with temperature is obvious. However, they have been detected on the same spectrum. In our measurements, we observed also two different gaps with different temperature dependencies on same spectrum [8]. In the end, we hope that we showed the existence of two energy scales in cuprates, particularly in YBCO and Bi2212.

## 5. ANALOGY BETWEEN THE SC IN UPd$_2$Al$_3$ AND IN CUPRATES

In this Section, we consider the analogy and difference between the SC in heavy fermion compound UPd$_2$Al$_3$ and the SC in hole-doped cuprates in frameworks of the MCS model.

The heavy fermion compound UPd$_2$Al$_3$ shows coexistence of both SC ($T_c$ = 2 K) and AF ordering ($T_N$ = 14 K) [47]. Quasiparticle charge carriers in UPd$_2$Al$_3$ have a heavy effective mass ($m^* \sim 100 m_0$) and can be considered as heavy magnetic polarons [10] or ferrons [48]. They couple to each other below $T_c$ by spin-fluctuations [47]. In fact, the SC scenario in UPd$_2$Al$_3$ is a very similar to a scenario for HTSC, proposed on earlier stage by Hizhnyakov and Sigmund [48].

The SC in cuprates coexist also with AF ordering. By analogy with the SC in heavy fermion compound UPd$_2$Al$_3$ and in frameworks of the MCS model, *fluctuating* 1D charge stripes in cuprates can be considered as "giant 1D magnetic polarons" which couple to each other below $T_c$ by spin fluctuations (see Fig. 1). On the other hand, these "giant 1D magnetic polarons" are not simple charge carriers at temperatures just above $T_c$ but carriers of preformed Cooper pairs. This is the main difference between the SC in heavy fermion compound UPd$_2$Al$_3$ and the SC in hole-doped cuprates. The 1D configuration of stripes is the key feature of the SC in cuprates. Spinons exist due to formation of 1D structure. If stripes could have 2D configuration, then, the $T_c$ in cooper-oxides could be as low as that in heavy fermion compound UPd$_2$Al$_3$. One has to take this fact into account in search of SC compounds with higher $T_c$.

The resonance peak observed by INS in cuprates [27] has been detected also in heavy fermion compound UPd$_2$Al$_3$ [49] in which the SC is mediated by spin fluctuation [47]. Thus, this experimental fact points out on the presence of the SC mediated by spin fluctuations in hole-doped cuprates.

## 6. DISCUSSION

In this Section, we discuss two issues. We begin from the fact that, in tunneling measurements on Tl2201 single crystals, performed by point-contact tunneling (PCT) technique along c-axis [36], it is not possible to observe the larger gap, the $\Delta_s$ (see Section 4). At the same time, tunneling spectroscopy on YBCO and Bi2212 shows clearly the presence of the $\Delta_s$. We know that the $\Delta_s$ is present in Tl2201 too [22], however, it doesn't show up in tunneling measurements. What is the reason for this? It is well known that the tunneling current probes a region of the order of the coherent length [15]. In the M(C)S scenarios, the coherent length perpendicular to stripes (or of magnetic polarons) is always larger than the spinon coherent length along charge stripes since $\Delta_s > \Delta_c$, always (see Fig. 2). The coherent length of spinons can be very short because spinons have no problems with the Coloumb repulsion. So, it seems that the reason, why the $\Delta_s$ can not be observed by tunneling spectroscopy along c-axis, is simply the geometry of one-layer cuprates. Tl2201 is one-layer compound, and magnetic interactions into $CuO_2$ planes are weaker than in double-layer compounds, especially, on the surface. Thus, the distance from the sample surface to the nearest $CuO_2$ plane carrying spinon pairs is larger than the coherent length of spinon pairs. It seems that the tunneling current can not reach spinon pairs. Tunneling measurements on one-layer NCCO single crystals [50], performed by STM along c-axis, show only the coherent gap too. PCT measurements on polycrystalline samples of $La_{1.85}Sr_{0.15}CuO_4$ [51] and $HgBa_2CuO_{4+x}$ [52] also show only the small energy scale, $2\Delta_c/k_BT_c \sim 4.8$ [53]. It seems that this is the case for all one-layer cuprates if tunneling measurements performed along c-axis. At the same time, in break-junction measurements on one-layer underdoped Bi2201 single crystals along $CuO_2$ planes, the $2\Delta_c/k_BT_c$ value can be the order of 20 [54]. It is already a sign of the $\Delta_s$ (see Fig. 2).

The second important issue which we would like to discuss is the subgap (4 -10 meV) observed in tunneling spectra of YBCO. Tunneling spectra measured in the center of the vortex core show a small gap of 5.5 meV [17], which we attributed to the spinon SC on chains, induced from outside the vortex core (see Section 3). Thus, it is possible to explain the presence of the subgap in spectra of YBCO in zero magnetic field by the SC on chains. However, a subgap with similar magnitude of 8-9 meV is often observed in tunneling spectra of Bi2212 [6,7], see, for example, the spectrum B shown in Fig. 7(b). There are no chains in Bi2212. We explained the presence of this subgap in spectra of Bi2212 by the presence of a g-wave SC mediated by spin fluctuations [6]. There are two reasons for this. First of all, there is no doubts that this subgap in Bi2212 has a magnetic origin [8]. Secondly, theoretically, the d-wave and a g-wave magnetically mediated SCs can coexist [55] with the ratio of gap magnitudes of, approximately, 2:1, respectively. So, it may be the right explanation. However, more research is needed to find out the exact origin of this subgap in Bi2212. If the subgap in Bi2212 is, indeed, the g-wave SC gap, then, in tunneling spectra of YBCO, there is the superposition of the g-wave gap and the SC gap on chains since they have similar magnitudes. This circumstance may explain the fact that, in YBCO, the zero bias conductance is *always* high in comparison of, for example, that in Bi2212.

All data presented in the present work show a good agreement with the M(C)S models. It seems, that the MCS model is more likely a SC scenario in hole-doped cuprates rather than the MS model. However, future investigations will answer this question more precisely. What is interesting that the PCS model for the electron doped NCCO cuprate is very plausible (see Section 4). If the PCS model is correct for the SC in NCCO, we predict that tunneling measurements on NCCO in the center of the vortex core at moderate $H$ will show also the presence of a large gap similar to the pseudogap observed in Bi2212 and YBCO [13].

## 7. CONCLUSIONS

We presented an experimental evidence in favor of the M(C)S models proposed earlier. We are not aware of the existence of another model of the SC in hole-doped cuprates which is capable to explain experimental data as the M(C)S models do. It seems that the PCS model is very likely scenario in electron-doped NCCO.

I thank A. N. Lavrov for sending Ref. 24 prior to publication, S. I. Vedeneev, A. G. M. Jansen, N. Miyakawa and R. Deltour for discussions. This work is supported by PAI 4/10.

**FIGURE CAPTIONS**

FIG. 1. Static picture of the stripe model [5], shown schematically. The temperature dependencies are shown also schematically.

FIG. 2. Idealized phase diagram of two SC gaps in Bi2212 [8]: straight line (spinon gap, $\Delta_s$) and parabolic line (coherent gap, $\Delta_c$). Inset: shapes of two SC gaps on the Fermi surface in Bi2212: black area (coherent gap having d-wave symmetry) and outlined area (spinon gap having entirely or partially a s-wave symmetry). The shapes of two gaps are shown schematically for the case of slightly overdoped Bi2212.

FIG. 3. Tunneling spectroscopy of Bi2212: (a) underdoped and (b) overdoped [13]. The upper and middle spectra in each frame were obtained at 4.2 K and 6 T between vortices and at the center of a vortex core, respectively. The lower spectra in each frame were measured above $T_c$ in zero magnetic field. They illustrate the striking resemblance of the vortex core pseudogap and normal state pseudogap [13]. The spectra have been shifted vertically for clarity.

FIG. 4. Tunneling spectra of two Bi2212 break junctions at 4.2 K in 0 and 20 T [15]. The orientation of the magnetic field is indicated. The spectra have been shifted vertically for clarity.

FIG. 5. Tunneling spectra for YBCO/Pb junction at 10 K in 0 T (solid line) and 8 T ($B \parallel$ c-axis) (dotted line) [16].

Fig. 6. Tunneling spectra of YBCO at 4.2 K: measured in zero magnetic field (upper curve) and average spectrum of spectra obtained at the center of a vortex core along a 5 nm path and in H = 6 T (lower curve) [17]. The upper spectrum has been shifted vertically for clarity.

FIG. 7. Tunneling spectra of: (a) Tl2201 with $T_c \sim 91$ K, measured in SIN (Au tip) junction at 4.2 K [36]; (b) Bi2212 with $T_c = 89.5$ K, measured in one SIS break junction at 14 K [7]; and (c) YBCO with $T_c = 89$ K (upper curve) and $T_c = 91$ K (lower curve), measured in SIN junctions at 10 K and 4.2 K, respectively [16,17]. The spectra B and C are obtained on *same* junction. The spectra B and E have been shifted vertically for clarity.

FIG. 8. Measured temperature dependencies of the quasiparticle DOS in Bi2212 [7,34,42,44] and YBCO [43] single crystals. The thick solid line corresponds to the BSC temperature dependence. One can clearly observe two types of $\Delta(T)$: above and below the BCS temperature dependence. The thin solid line and dashed lines are guides to the eye.

Fig. 9. Two temperature dependencies of the quasiparticle DOS, measured simultaneously on a Bi2212 single crystal [38]. The solid line corresponds to the BSC temperature dependence. The dashed lines are guides to the eye.

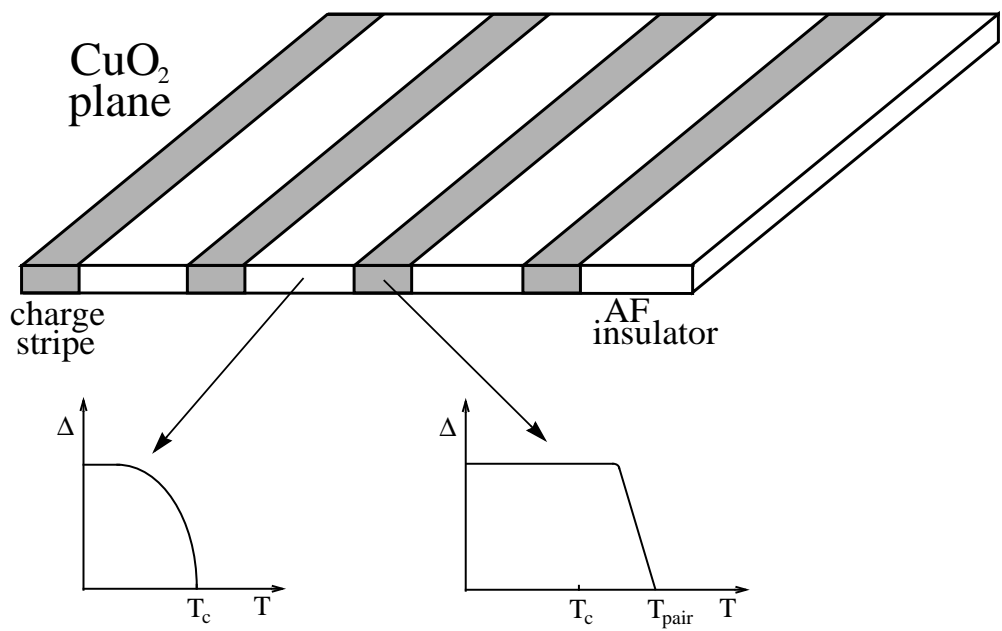

FIG. 1

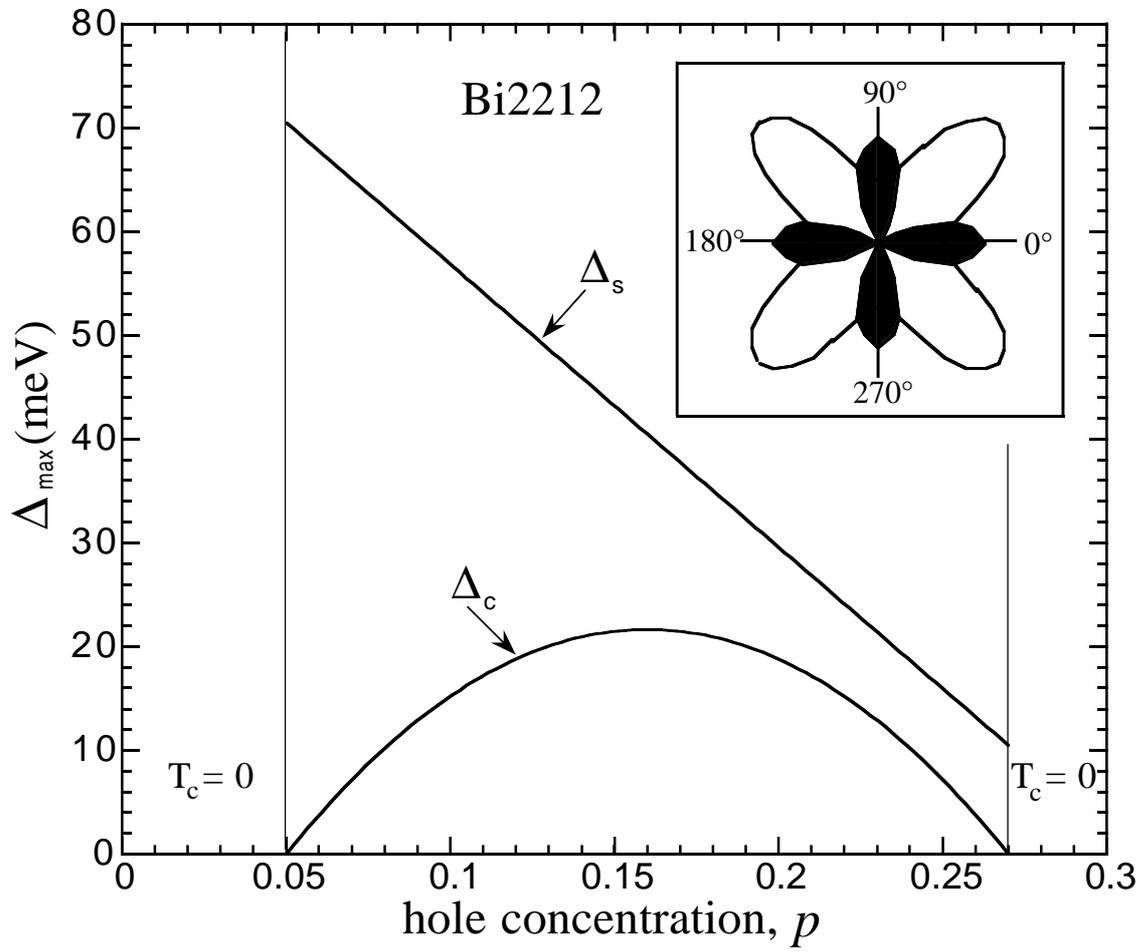

FIG. 2

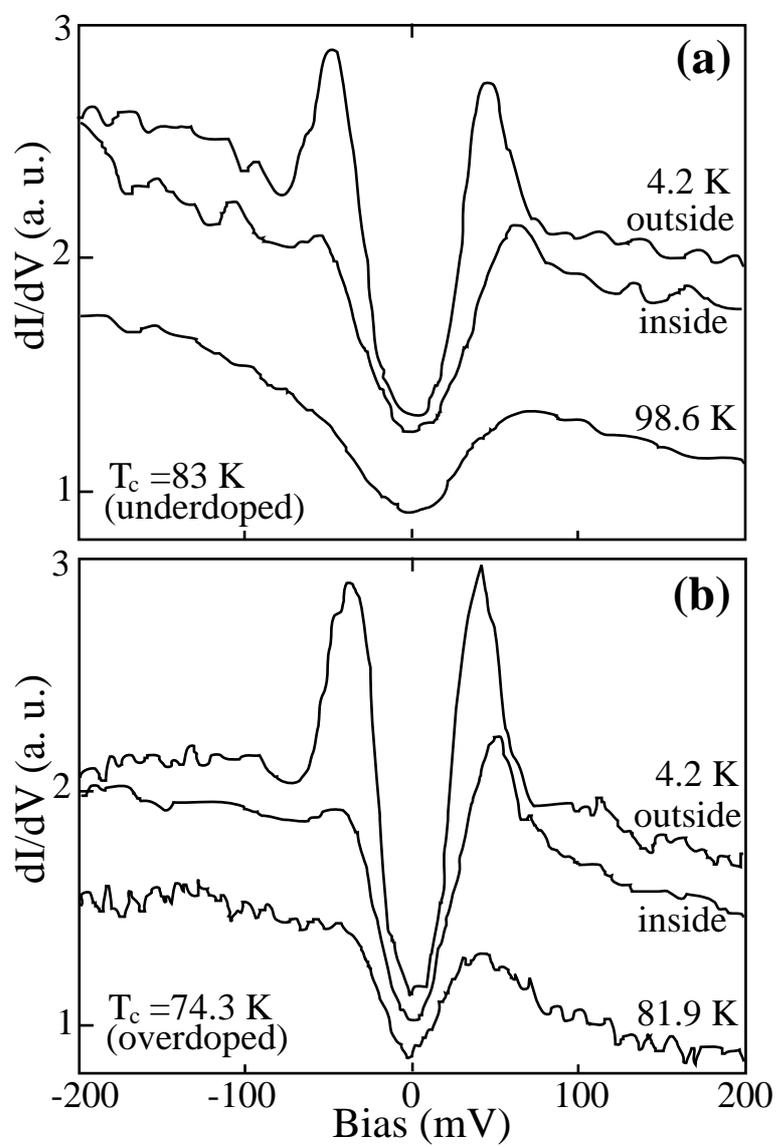

FIG. 3

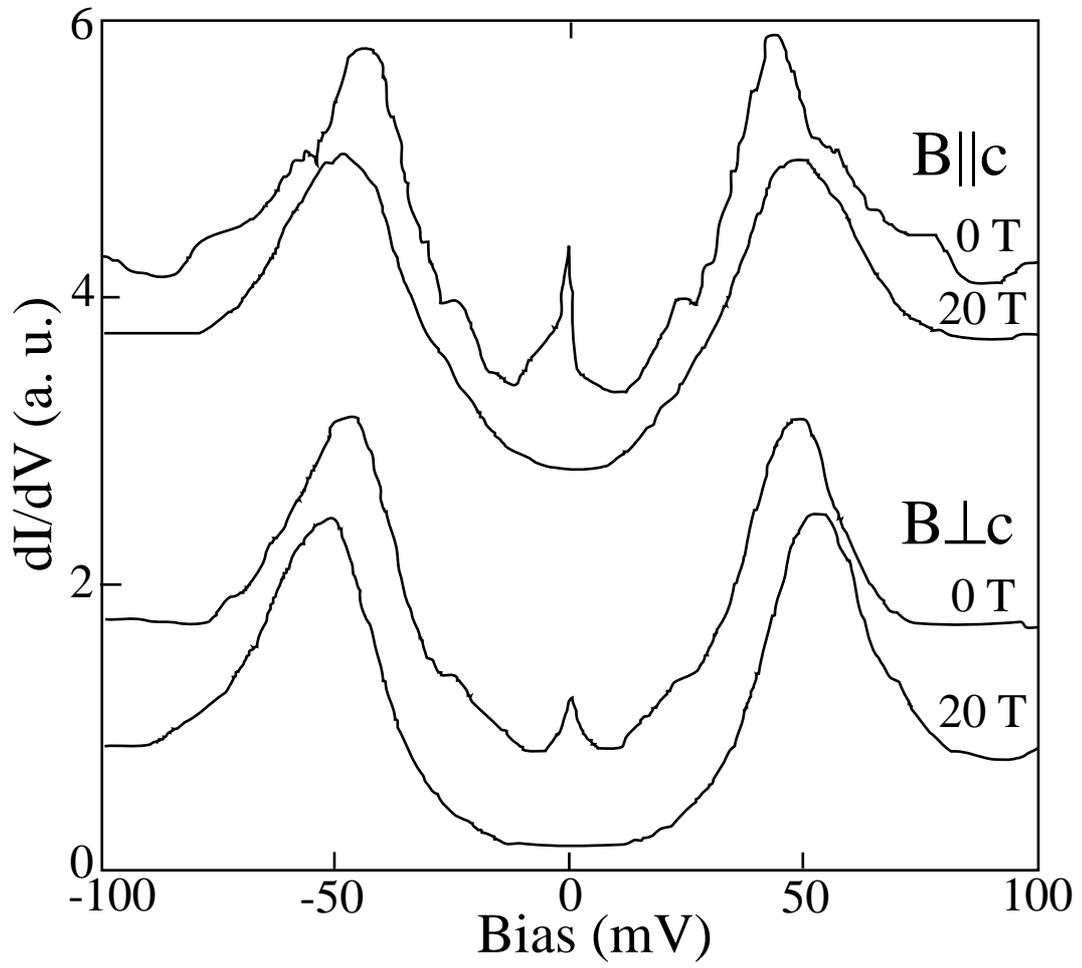

FIG. 4

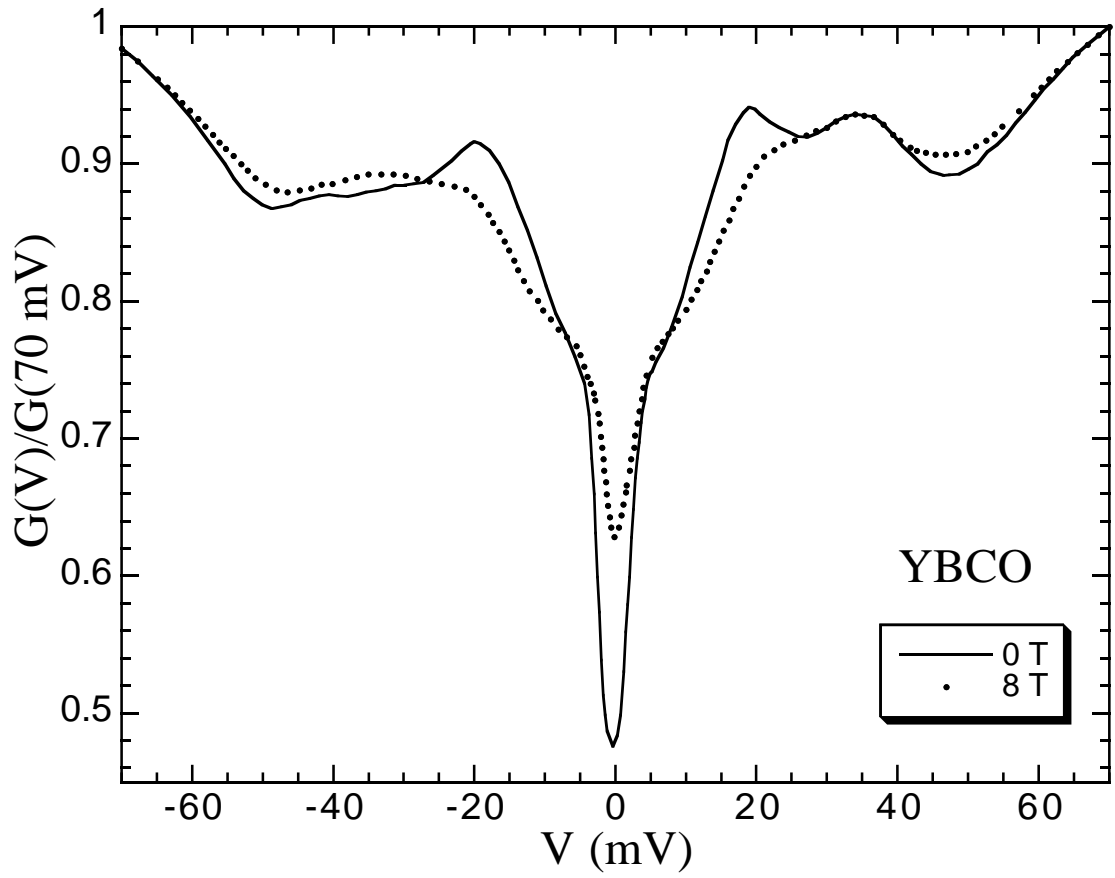

FIG. 5

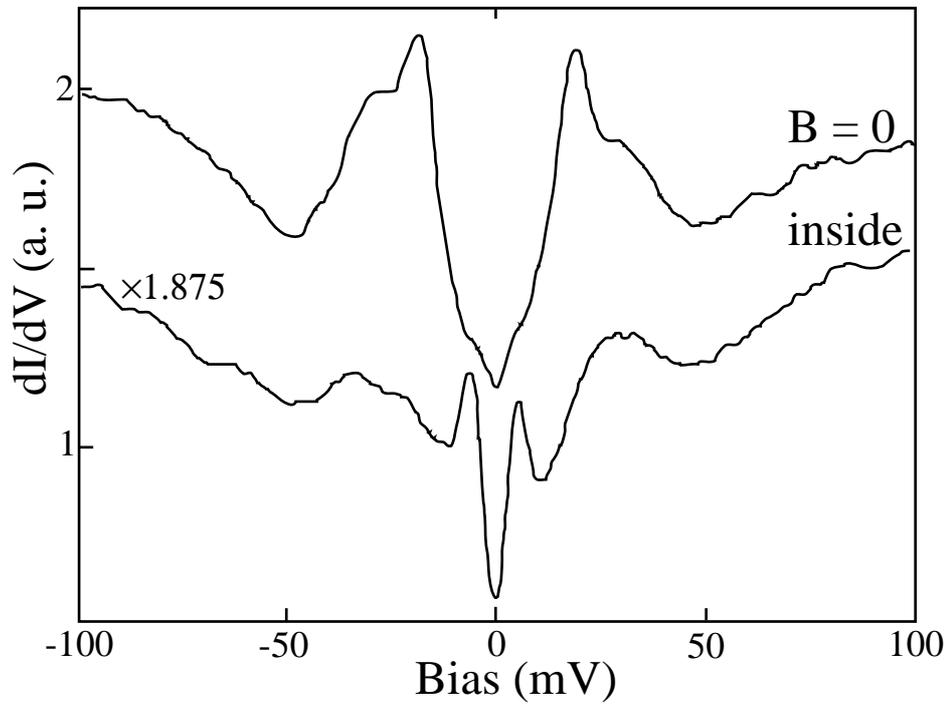

FIG. 6

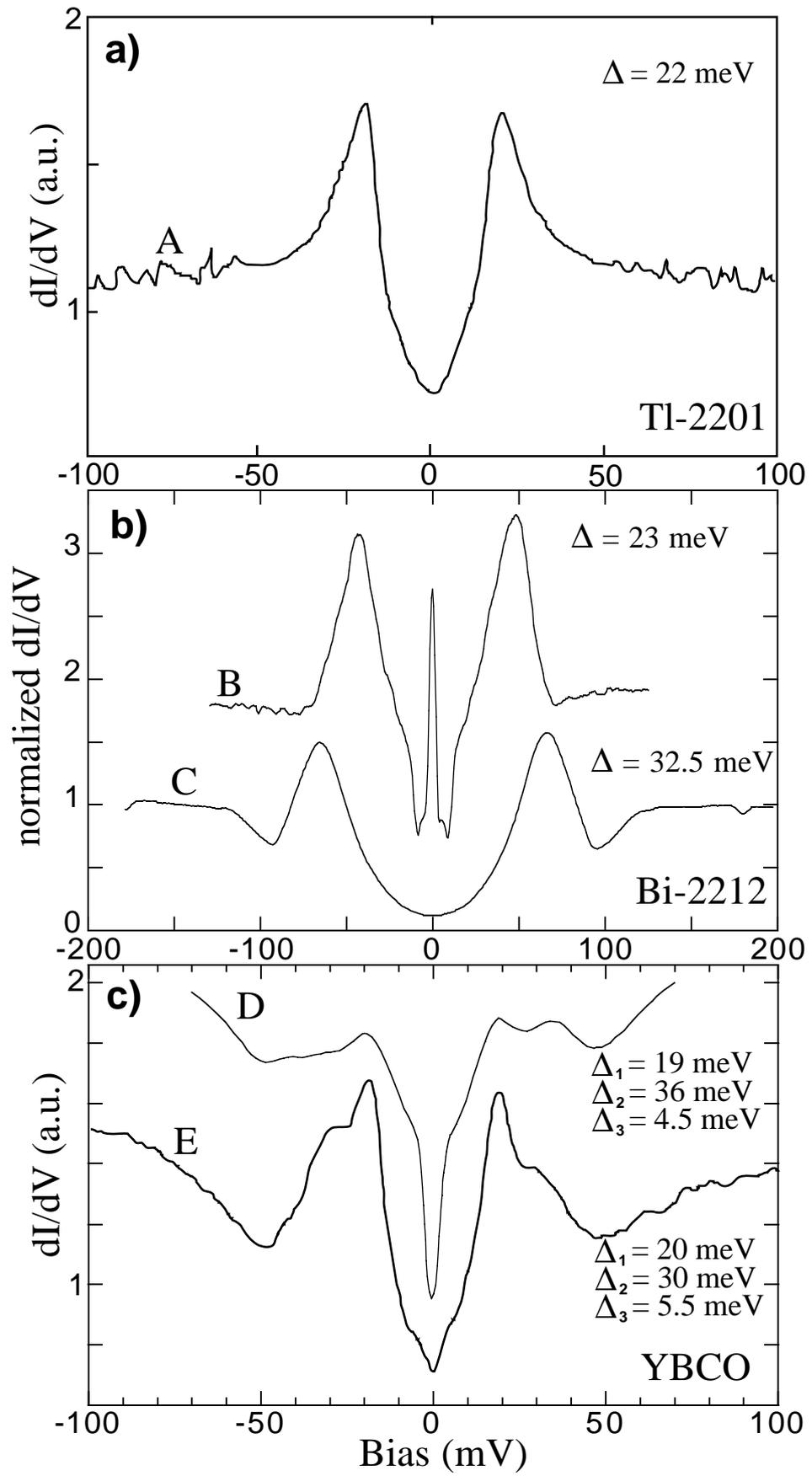

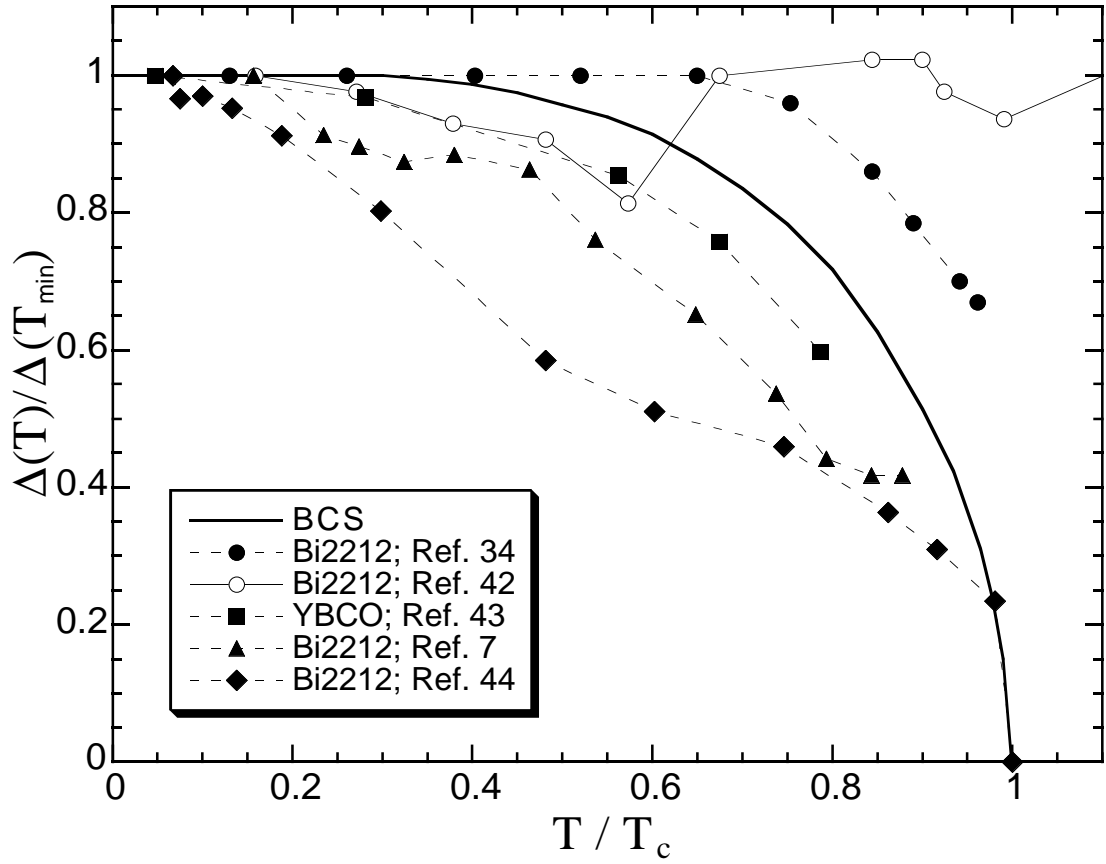

FIG. 8

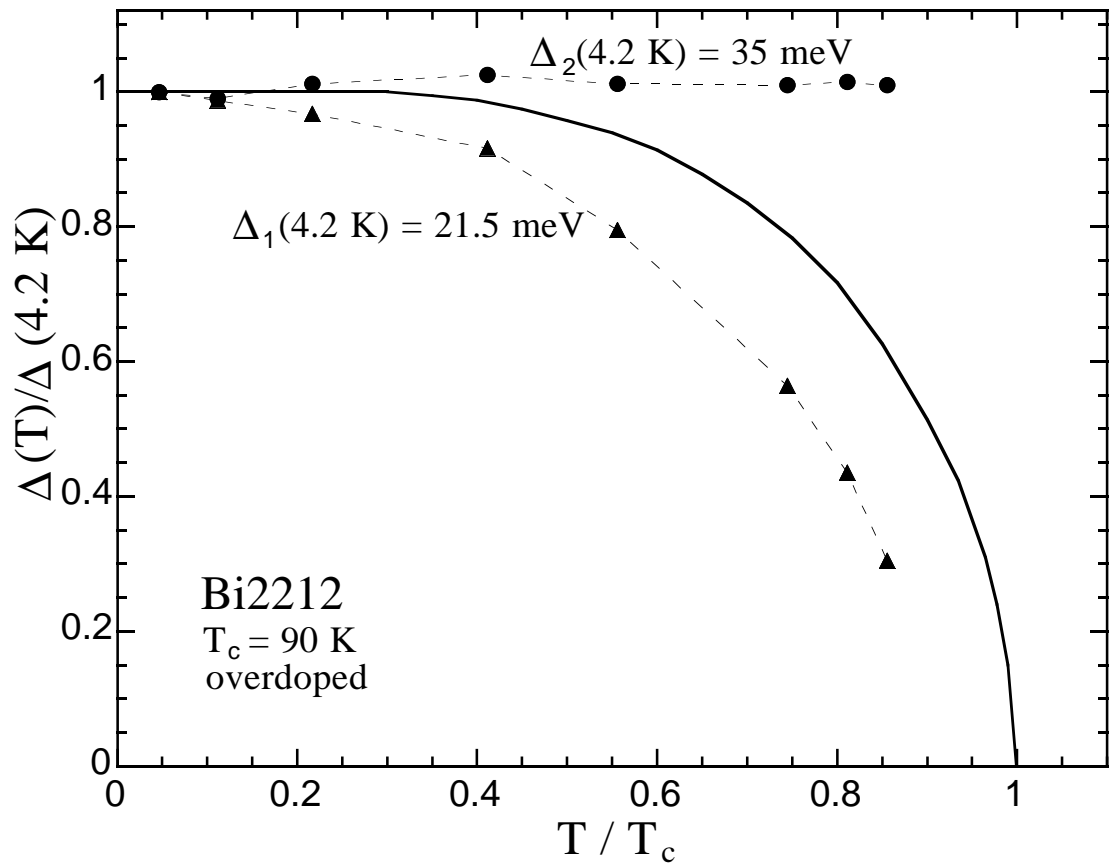

FIG. 9